\documentclass[prb,showpacs,twocolumn,superbib,floats,groupedaddress]{revtex4-1}
\usepackage{graphicx}
\usepackage{amsmath}
\usepackage{bm}
\usepackage{natbib}
\usepackage{multirow}
\usepackage{longtable}

\begin{document}

\title{Phonons in single and few-layer MoS$_2$ and WS$_2$}

\author{A. Molina-S\'{a}nchez}
\email[corresponding author: ]{alejandro.molina@isen.iemn.univ-lille1.fr}
\author{L. Wirtz}
\affiliation{Institute for Electronics, Microelectronics, and 
Nanotechnology (IEMN), CNRS-UMR 8520, Department ISEN, B.P. 60069, 59652 
Villeneuve d'Ascq, France}

\date{\today}

\begin{abstract}

We report \textit{ab-initio} calculations of the phonon dispersion relations
of the single-layer and bulk dichalcogenides MoS$_2$ and
WS$_2$. We explore in detail the behavior of the Raman active modes, 
$A_{1g}$ and $E_{2g}^1$ as a function of the number of layers. In
agreement with recent Raman spectroscopy measurements 
[C. Lee et. al., ACS Nano \textbf{4}, 2695 (2010)] we find that
the $A_{1g}$ mode increases in frequency with increasing
layer number while the $E_{2g}^1$ mode decreases. We explain this decrease
by an enhancement of the dielectric screening of the long-range Coulomb 
interaction between the effective charges with growing number of layers. 
This decrease in the long-range part over-compensates the increase of 
the short-range interaction due to the weak inter-layer interaction.

\end{abstract}

\maketitle

\section{Introduction}

Recently, the mechanical exfoliation technique, applied to
layered materials, has lead to the fabrication of new 
bidimensional systems, i.e., atomically thin layers.\cite{Novoselov2005a}
Graphene,\cite{Novoselov2004} a planar sheet of carbon atoms arranged in 
a hexagonal lattice, is the most famous bidimensional material and
exhibits intriguing physical properties not found in 
its bulk counterpart graphite.\cite{Wallace1947,Geim2007} However, 
the absence of a bandgap makes its use in electronic devices (transistors) 
difficult. Several strategies were proposed to overcome this setback by opening
a gap: quantum confinement in nanoribbons,\cite{son2006} 
deposition of a graphene monolayer on boron nitride,\cite{giovannetti2007}
applying an electric field in bilayer graphene,\cite{Zhang2009a} etc.
Nevertheless, the experimental realization of a bandgap larger than
400 meV remains a challenge,\cite{Han2007} besides of 
a deterioration of other graphene properties, in particular the high mobility. 
Therefore, the fabrication of atomically thin sheets
of other materials, with a finite bandgap, appears as the natural strategy
in the search of materials for a new generation of electronic 
devices.

In recent experiments, molybdenum disulfide (MoS$_2$), 
an indirect semiconductor of bandgap 1.29 eV in its bulk phase, has
exhibited a direct bandgap of 1.75 eV in its single-layer 
phase, together with an enhancement of the luminescence
quantum yield in comparison with the MoS$_2$ bulk.\cite{Splendiani2010,Mak2010}
Moreover, Radisavljevic \textit{et. al.}\cite{RadisavljevicB.2011} 
have demonstrated suitable properties of a 
single-layer MoS$_2$-based transistor,
like a room-temperature electron
mobility close to that of graphene nanoribbons and a high on/off ratio. 
Therefore, single-layer MoS$_2$ has become an 
appealing material in the area of
optoelectronic devices, being an alternative and/or complement
to graphene. From other layered compounds such as WS$_2$, 
MoS$_2$, BN,\dots monolayers can be produced by (liquid) exfoliation as 
well.\cite{Coleman2011}
Moreover, MoS$_2$ and other layered materials
are also interesting due to change of certain properties with respect 
to its bulk counterparts. Finally, their topology facilitates
the chemical identification atom-by-atom.\cite{Krivanek2010}

Recent Raman spectroscopy measurements of MoS$_2$ single and multi-layers
have revealed unexpected trends of the vibrational 
properties when the number of layers changes.
Lee \textit{et. al.}\cite{Lee2010} reported
a decrease in frequency of the optical $E_{2g}^1$ phonon mode with 
increasing number of layers. This is consistent with the finding that in
bulk MoS$_2$ the infrared active $E_{1u}$ mode (where neighboring layers are 
vibrating in-phase) is slightly lower in frequency than the Raman active $E_{2g}^1$
mode (where neighboring layers are vibrating with a phase shift of $\pi$).\cite{Wieting1971} But both 
observations contradict the naive expectation
that the week inter-layer forces should increase the effective restoring 
forces acting on atoms. One would thus rather expect a slight increase
of the frequency of the Raman active mode with respect to the IR active mode
\cite{Ghosh1983} and, accordingly, an increase of the frequency of the bulk 
Raman active mode with respect to the corresponding single-layer mode. As
a plausible explanation of this anomalous trend the long range
Coulomb interaction was mentioned.\cite{Lee2010}
The purpose of our article is the clarification of this issue by a detailed
ab-initio study of the inter-atomic force constants, separating the short-range
and the long range Coulomb parts. We show in the following,
that the anomalous trend in the $E_{2g}^1$ mode frequency is caused by the 
dielectric screening of the long range Coulomb forces in bulk MoS$_2$.
At the same time, we present a full ab-initio study of the phonon dispersion
relations of the phonon dispersion relations of single-layer and of bulk 
MoS$_2$ and the intimately related material WS$_2$. (Tungsten is in the 
same column of the periodic system as Molybdenum). Apart from a study of the
vibrational stability of MoS$_2$ nanoribbons,\cite{Ataca2011}
a fully comprehensive \textit{ab initio} study of the 
vibrational properties of these materials is still absent in the literature.

In Section II we present the methods for the calculation of the phonon 
dispersions and the analysis of the force constants.
In section III we discuss the phonon dispersion relations of MoS$_2$ and
WS$_2$ single layers and bulk and compare with experimental data.
In section IV, we present the calculated results for the evolution of the Raman 
active phonon modes as a function of the number of layers and give an 
explanation in terms of the short-range and long-range contributions to the
force constants.

\section{Calculation Methods}
\label{theory}

\begin{figure}
\begin{center}
\includegraphics[width=7.6 cm]{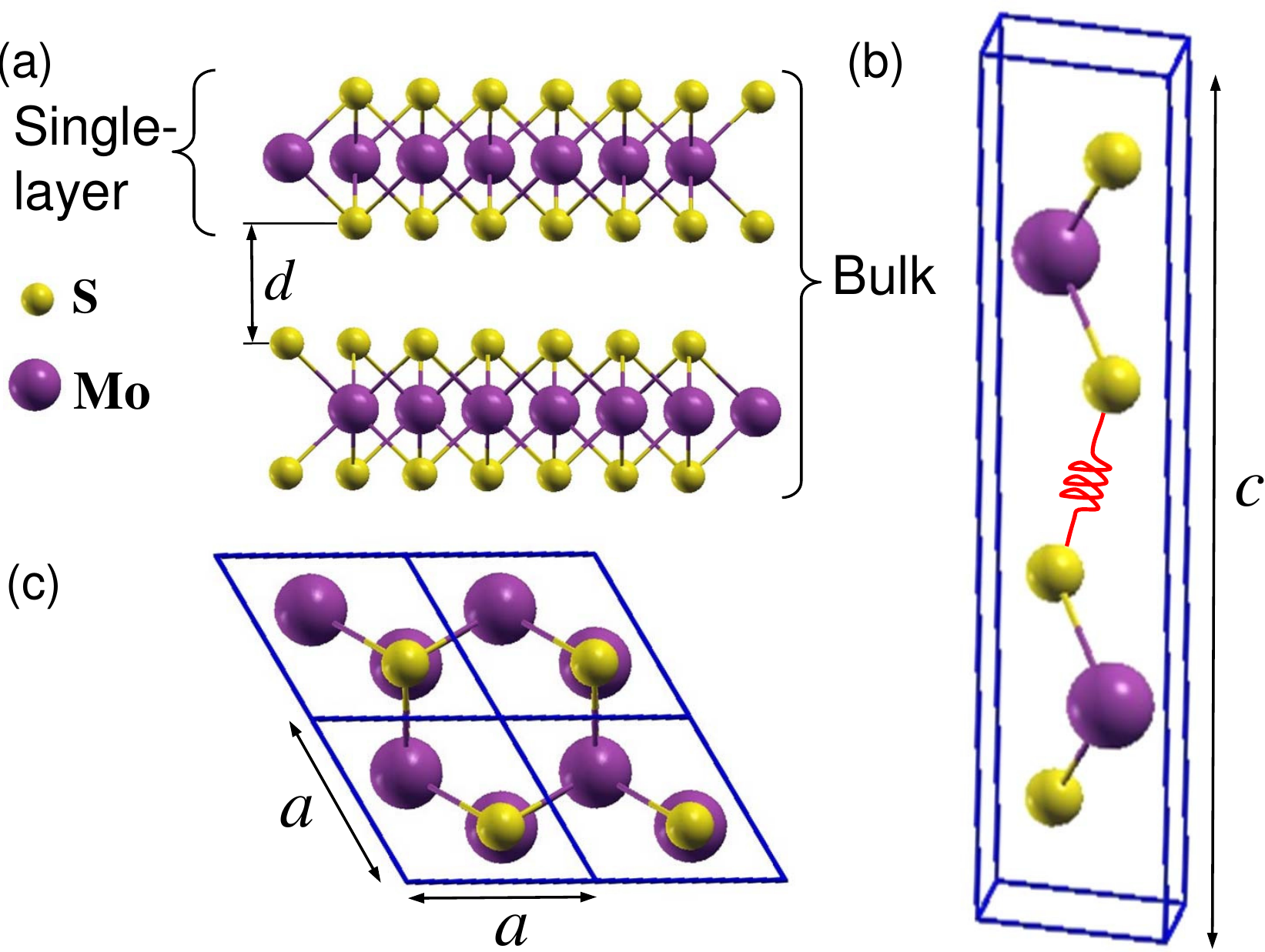}
\caption{(a) MoS$_2$ bulk and single-layer. The interlayer distance
is denoted by $d$. (b) and (c) Slide and top view 
of the MoS$_2$ bulk unit cell (analogously 
for WS$_2$) unit cell. The primitive vectors are $\bm{a}=(a,0,0)$, 
$\bm{b}=(a/2,\sqrt{3}a/2,0)$, and $\bm{c}=(0,0,c)$. The weak layer interaction
is indicated by a spring.}
\label{unitcell}
\end{center}
\end{figure}

MoS$_2$ and WS$_2$ belong to the dichalcogenide family  of materials,
built up of weakly (van der Waals) bonded S-Mo-S single-layers as 
shown in Fig.~\ref{unitcell}. Each one of these single-layers  
consists of two hexagonal planes of S atoms and an intercalated 
hexagonal plane of Mo atoms bound with the sulfur atoms in 
a trigonal prismatic arrangement. The symmetry space group
of bulk MoS$_2$ is $P3m1$ (point group $D_{6h}$). The space
group of the single layer is $P6m2$ (point group $D_{3h}$). Consequently, 
systems with even number of layers belong to the space group
$P3m1$ (with inversion symmetry), and systems with odd number of layers 
to the $P6m2$ space group (without inversion symmetry). 

The phonon calculations begins with the determination 
of the equilibrium geometry (i.e. the relative
atomic positions in the unit cell which yield zero forces 
and the lattice constants which leads to a zero-stress tensor). The
calculations have been done with
density functional theory (DFT) as implemented in
the open-source code ABINIT,\cite{abinit} within 
the local density approximation (LDA).\cite{Kohn1965} 
We use Hartwigsen-Goedecker-Hutter pseudopotentials\cite{hgh}
(including the semi-core electrons as valence electrons in the 
case of Mo and W) and a plane-wave cutoff at 60 Ha.
The first Brillouin zone is sampled with a $12\times 12\times 4$ 
Monkhorst-Pack grid for bulk and $12\times 12\times 1$ for 
single and few-layers systems.

The optimized lattice parameters are shown in table~\ref{lattice}. The 
experimental lattice parameters of MoS$_{2}$ are $a=3.15$ and $c=12.3$ \AA.\cite{Wakabayashi1975} In the case
of WS$_2$ they are $a=3.153$ and $c=12.323$ \AA.\cite{Schutte1987}
Our LDA calculations underestimate them by
0.7 \% and 2.1 \%, respectively. 
The slight underestimations of the in-plane lattice constant is a common
feature of the LDA which tends to overestimate the strength of covalent
bonds. For the single-layer, we checked the influence of the exchange-correlation
potential on the geometry and the phonon dispersion by performing
calculations within the Generalized Gradient Approximation (GGA).\cite{burke} For
the single-layer of both MoS$_2$ and WS$_2$, we obtain a lattice constant
$a=3.18$ \AA, 1.76 \% larger than the LDA value and 0.96 \% larger than 
the experimental (bulk) value. Correspondingly, the phonon frequencies are reduced
by an almost constant factor of 1.04 \% throughout the whole phonon dispersion.

The correct description of the $c$ parameter is less evident
because the LDA (and other semi-local functionals) completely neglect
the van der Waals component of the inter-layer-interaction. At the same time,
however, the LDA strongly overestimates the (weak) covalent part of the
inter-layer bonding. Thus, the LDA has quite successfully reproduced the geometry
and also given reasonable results for layer phonon modes
of different layered materials such as graphite\cite{kresse} and hexagonal 
boron nitride (hBN),\cite{kern} as well as of single-layers of hBN\cite{serrano}
and graphene\cite{allard} on a Ni(111) surface. We thus expect (and the
obtained value for $c$ supports this expectation) that
the LDA works reasonably well for the inter-layer phonons of the
MoS$_2$. We 
note that a physically correct description of the equilibrium geometry, and 
the potential energy surface around, would require 
the proper treatment of van der Waals contribution, 
e.g. on the level of the random-phase approximation as it has been 
done for bulk hBN\cite{Marini2006} and for graphite.\cite{lebegue} Alternatively,
non-local functionals that are optimized for the description of the
van der Waals interaction\cite{Rydberg2003} could be used.
Both approaches are, however, out of the scope of the present work.

For the calculations of single-layer and few-layers systems, we have used
a periodic supercell, leaving enough distance between adjacent
sheets. For instance, we use $c=13.25$ Angstroms in the case of a 
single-layer. The remaining interlayer interaction
has negligible effects on the phonon frequencies. All the 
results show a slight reduction of the in-plane lattice constant
together with a slight stretching of the vertical 
distances, with the total effect of a smaller Mo-S bond length
for decreasing number of layers, being 2.382~\AA~for single-layer and 2.384~\AA~for
bulk MoS$_2$.

\begin{table}
\begin{tabular}{lccc}
\hline
\hline
\multicolumn{4}{c}{Lattice constants (\AA)} \\
\hline
MoS$_2$ & 1-layer & 2-layer & Bulk   \\
$a$     &  3.125  &  3.126  & 3.127  \\
$c$     &   -     &   -     &12.066  \\
\hline
WS$_2$  & 1-layer & 2-layer & Bulk   \\
$a$     & 3.125   & 3.125   &  3.126 \\
$c$     &   -     &   -     & 12.145 \\
\hline
\hline
\end{tabular}
\caption{Equilibrium lattice parameters of 
MoS$_2$ and WS$_2$ obtained in this work.}
\label{lattice}
\end{table}


Once the equilibrium geometry has been obtained, the phonon
frequencies $\omega$ can be calculated. These phonon 
frequencies are the solution of the secular equation

\begin{equation}
\left| \frac{1}{\sqrt{M_IM_J}}\tilde{C}_{I\alpha,J\beta}(\bm{q}) 
- \omega^2(\bm{q})\right| = 0,
\label{secular}
\end{equation}
where $\bm{q}$ is the phonon wave-vector, and $M_I$ 
and $M_J$ are the atomic masses of 
atoms $I$ and $J$. The dynamical matrix is then defined as

\begin{equation}
\tilde{C}_{I\alpha,J\beta}(\bm{q})
=\frac{\partial^2 E}{\partial u^{*\alpha}_I(\bm{q})
\partial u^{\beta}_J(\bm{q})},
\label{sec_der}
\end{equation} 
where $u^{\alpha}_I(\bm{q})$ denotes the displacement 
of atom $I$ in direction $\alpha$. The second
derivative of the energy in Eq.~\ref{sec_der} corresponds 
to the change of the force acting on atom $I$
in direction $\alpha$ with respect 
to a displacement of atom $J$ in direction $\beta$:\cite{Baroni2001}

\begin{equation}
\tilde{C}_{I\alpha,J\beta}(\bm{q}) 
= -\frac{\partial F^{\alpha}_{I}}{\partial u^{\beta}_J(\bm{q})}.
\label{force}
\end{equation} 

The Fourier transform of the $\bm{q}$-dependent matrix leads to
the real space atomic force constant matrix $C_{I\alpha,J\beta}(\bm{R}_{IJ})$,
where $\bm{R}_{IJ}$ is the vector that joins atoms $I$ and $J$. Thus,
$C_{I\alpha,J\beta}<0\quad(>0)$ means a 
binding (anti-binding) force in direction $\alpha$ acting on atom $I$ when atom 
J is displaced in direction $\beta$. It is worth to mention that the 
diagonal term in the atom index, $C_{I\alpha,I\beta}$, corresponds, according
to Newton's third law, to the total
force exerted in the $\alpha$-direction on the atom $I$, when the displacements
of the atoms $J$ in the $\beta$-direction is are set to one:\cite{Bruesch1982}

\begin{equation}
C_{I\alpha,I\beta}(\bm{0}) = \sum_{J\neq I}^{\infty} \frac{\partial F^{\alpha}_{I}}{\partial u^{\beta}_J}.
\label{selfforce}
\end{equation}

This term is always positive (unless the crystal is unstable) and in the 
following we refer to it as self-interaction. 
Eq.~(\ref{selfforce}) demonstrates the contribution
of many atoms to the self-interaction. One can distinguish two
contributions, the short range part (which is mainly due to covalent bonding 
to the close neighbors) and the Ewald or long
range part\cite{Gonze1997} (due to the Coulomb forces between the effective 
charges). This distinction will be helpful to interpret the evolution of the
self interaction for varying layer thickness and to understand the unexpected
trends of the phonon frequencies (section \ref{secthick}).

For the calculation of the dynamical matrix we have used 
density functional perturbation theory (DFPT)\cite{Baroni2001} where
atomic displacements are taken
as a perturbation potential and the resulting changes in 
electron density and energy are calculated
self-consistently through a system of Kohn-Sham 
like equations. Within this approach the phonon frequency
can be obtained for arbitrary $\bm{q}$, with 
calculation only in a single unit-cell. 


Since MoS$_2$ and WS$_2$ are slightly polar 
materials, certain IR active phonon modes at $\Gamma$ give
rise to a macroscopic electric field. This electric fields affects
the longitudinal optical (LO) phonons in the limit $\bm{q}\rightarrow0$, 
breaking the degeneracy of the LO mode with the transversal optical (TO) 
mode.\cite{CardonaBook} Thus, in bulk MoS$_2$ and WoS$_2$, the nonanalytic 
part of the dynamical matrix (which contains the effective charges and the 
dielectric tensor) must be calculated in order to obtain  
the correct frequencies at the Brillouin zone-center.\cite{Giannozzi1991}
The LO-TO splitting for the $E_{1u}$ mode has the value of 2.8 cm$^{-1}$. 
In the case of a single or few-layers system, this effect is even smaller.

\section{Phonon Dispersions}

\subsection{MoS$_2$}

\begin{figure}
\begin{center}
\includegraphics[width=7.6 cm]{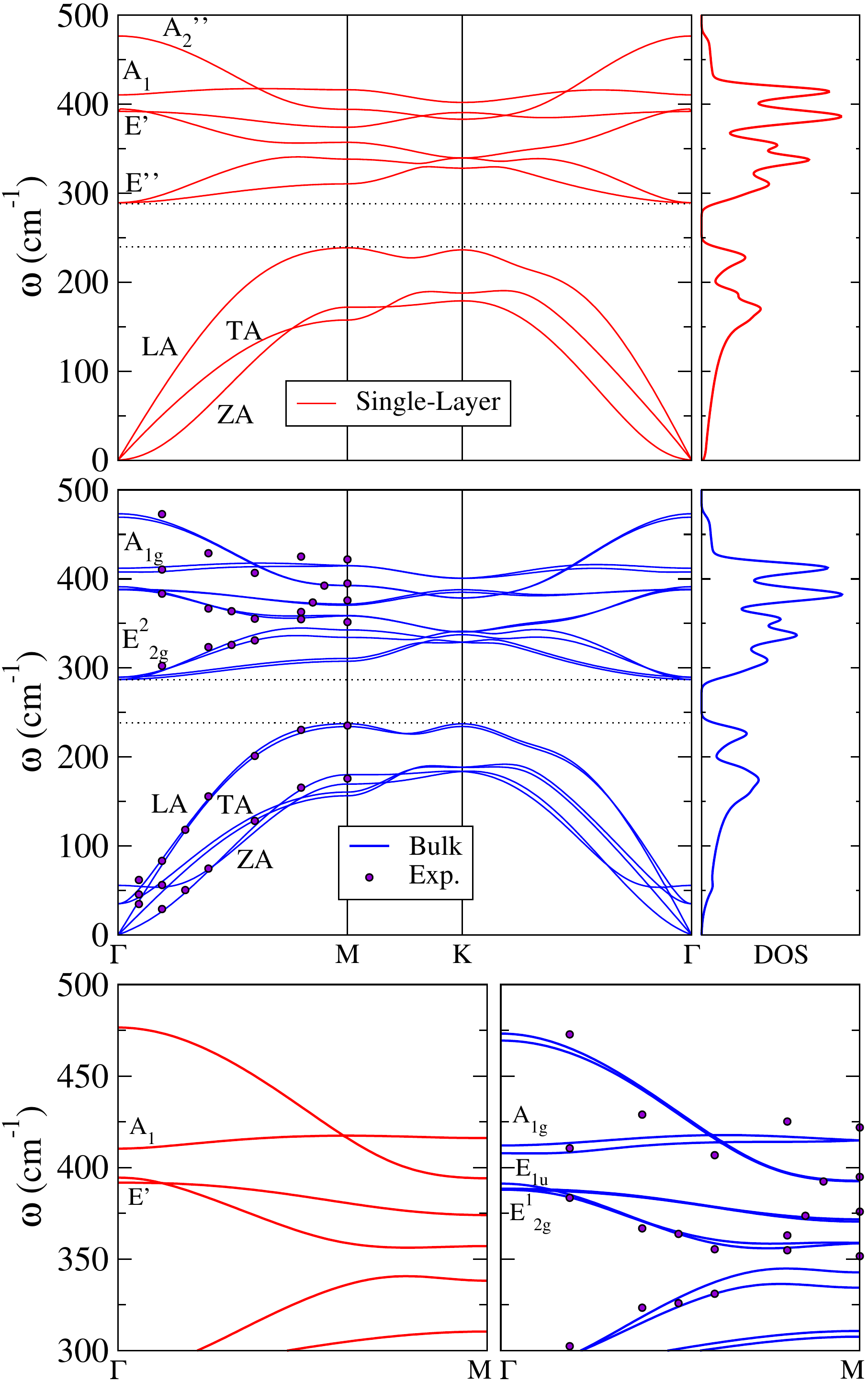}
\caption{Phonon dispersion curves and density of states of 
single-layer and bulk MoS$_2$. Points are experimental data 
extracted from Ref.~\onlinecite{Wakabayashi1975}. Low panel, inset of
the phonon branches in the region of the $E_{2g}^1$ and $A_{1g}$ modes.}
\label{mos2_phon}
\end{center}
\end{figure}

We start our analysis of the vibrational properties with the
description of the general features of the phonon dispersions of bulk
and single-layer MoS$_2$, shown in Figure~\ref{mos2_phon}.
We have also depicted the experimental data obtained with 
neutron inelastic scattering spectroscopy\cite{Wakabayashi1975}. 
The overall agreement between theory and experiment is good, even for the
inter-layer modes. 
This confirms our expectation that the LDA describes reasonably well
the inter-layer interaction (even though not describing the proper physics
of the inter-layer forces).

The bulk phonon dispersion has three acoustic modes. Those that
vibrate in-plane (longitudinal acoustic, LA, and transverse acoustic, TA) 
have a linear dispersion and higher energy than the out-of-plane acoustic (ZA) 
mode. The latter displays a $q^2$-dependence analogously 
to that of the ZA mode in graphene (which is a consequence of the
point-group symmetry\cite{Saito1998}). The low frequency optical modes
are found at 35.2 and 57.7 cm$^{-1}$ and correspond to rigid-layer shear/vertical 
motion, respectively (in analogy with the low frequency optical modes
in graphite\cite{Wirtz2004}). When the wave number
$\bm{q}$ increases, the acoustic and low frequency optical branches
almost match. It is worth to mention the absence
of degeneracies at the high symmetry points $M$ and $K$ and the two
crossings of the LA and TA branches just before and after the $M$ point. 

\begin{figure}
\begin{center}
\includegraphics[width=7.6 cm]{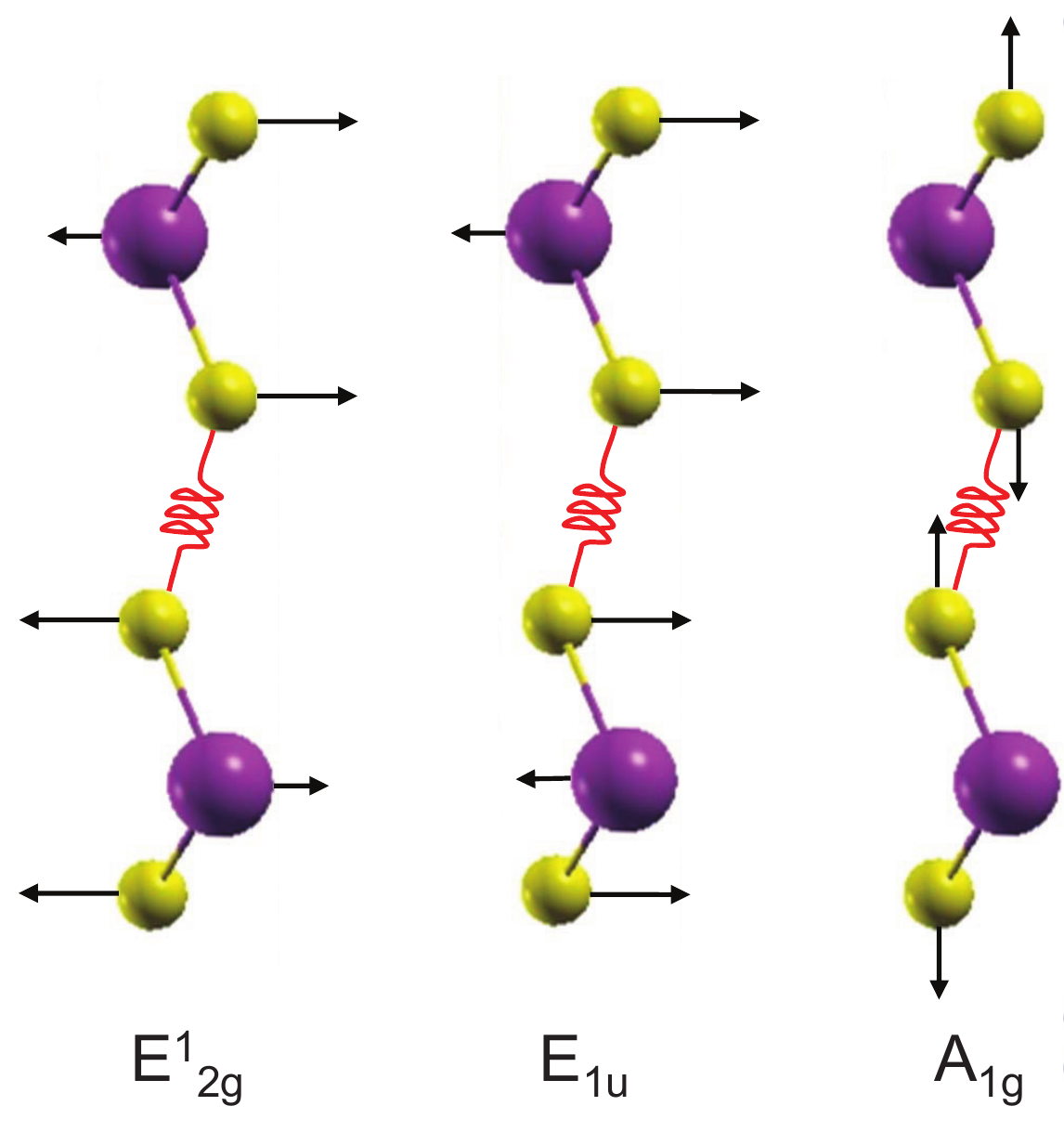}
\caption{Phonon modes in-plane $E^1_{2g}$, $E_{1u}$, and the 
out-of-plane phonon mode $A_{1g}$, for the 
bulk Mo$S_2$ (analogously for WS$_2$).}
\label{modes_MoS2}
\end{center}
\end{figure}

The high frequency 
optical modes are separated from the low frequency modes by a gap of 49 cm$^{-1}$.
We have drawn in Fig.~\ref{modes_MoS2} the atomic displacements of the Raman 
active modes ($E_{2g}^1$ and $A_{1g}$) and the infrared active mode $E_{1u}$. 
The Raman active modes are also 
indicated in the phonon dispersion of Fig.~\ref{mos2_phon}. 
The in-plane modes 
$E_{2g}^1$ and $E_{1u}$ are slightly split in energy (by 3 cm$^{-1}$).
This is known as Davydov splitting and, for MoS$_2$, the experimental value
is 1 cm$^{-1}$.\cite{Wieting1971} However, 
the finding that the $E_{1u}$ frequency is larger than that of the $E_{2g}^1$
mode contradicts \textit{a priori} what one would expect from the
weak inter-layer interaction: As one can see in Fig.~\ref{modes_MoS2}, for 
the mode $E_{2g}^1$, the sulfur atoms of different layers move in opposite
direction and thus the additional ``spring'' between sulfur atoms of neighboring
layers should increase the frequency of the mode $E_{2g}^1$ with respect
of that of $E_{1u}$ mode where sulfur atoms of neighboring sheets are moving
in phase and thus the additional ``spring'' has no effect. 
The semi-empirical model of Ref.~\onlinecite{Ghosh1983}
takes this consideration into account, and obtained indeed
that $\omega_{E_{2g}^1}>\omega_{E_{1u}}$ while experiments\cite{Wieting1971,Ghosh1983} demonstrate the opposite behavior. 
Our \textit{ab-initio} calculations match the experimental results
which indicates that other causes beyond the weak inter-layer interaction
are present in the system. We will analyze this
feature in the next Section with the aid of the atomic force constants. 

\begin{table}
\begin{tabular}{ccccccc}
\hline\hline
$D_{3h} $                   & $D_{6h}$   & Character & Direction & Atoms   & \multicolumn{2}{c}{$\omega_{MoS_{2}}$ (cm$^{-1}$)} \\
\hline
\multirow{2}{*}{$A_2$}      & $A_{2u}$   & Acoustic  & Out-of-plane & Mo+S & 0.0 & 0.0    \\
                            & $B^2_{2g}$ & Inactive  & Out-of-plane & Mo+S & -   & 55.7    \\
\hline
-                           & $E_{2g}^2$ & Raman  & In-plane & Mo+S & - & 35.2  \\
\hline
\multirow{2}{*}{$A_1$}      & $A_{1g}$   & Raman     & Out-of-plane & S    & \multirow{2}{*}{410.3}& 412.0 \\
                            & $B_{1u}$   & Inactive  & Out-of-plane & S    &                       & 407.8 \\
\hline
\multirow{2}{*}{$A_2^{''}$} & $A_{2u}$   & Infrared ($\bm{E}||\bm{c}$)  & Out-of-plane & Mo+S & \multirow{2}{*}{476.0} & 469.4 \\
                            & $B^1_{2g}$ & Inactive  & Out-of-plane & Mo+S                    &                        & 473.2 \\
\hline
\multirow{2}{*}{$E'$}       & $E_{2g}^1$ & Raman  & In-plane & Mo+S                  & \multirow{2}{*}{391.7} & 387.8 \\
                            & $E_{1u}$   & Infrared ($\bm{E}\bot\bm{c}$)  & In-plane & Mo+S &                 & 391.2 \\
\hline
\multirow{2}{*}{$E''$}      & $E_{1g}$   & Raman     & In-plane & S & \multirow{2}{*}{289.2} & 288.7 \\
                            & $E_{2u}$   & Inactive  & In-plane & S &                        & 287.1 \\
\hline
\hline
\end{tabular}
\caption{Relevant phonon symmetry representations 
of single-layer (point group $D_{3h}$) and
bulk (point group $D_{6h}$) MoS$_2$ (inspired in Table II of 
Ref.~\onlinecite{Wieting1971}). Direction
out-of-plane (in-plane) is 
parallel (perpendicular) to the $\bm{c}$ vector of the unit
cell, respectively. Phonon frequencies are the calculated values
of this work.}
\label{symmetry}
\end{table}

We now turn to analyze the single-layer phonon dispersion, shown in 
Fig.~\ref{mos2_phon}. The symmetry is reduced from $D_{6h}$ to $D_{3h}$ 
and there is no longer a center of inversion as in the bulk. Therefore,
the phonon mode labels at $\Gamma$ must be changed accordingly. The number of
phonon branches is reduced to nine. 
Table~\ref{symmetry} shows the most relevant 
MoS$_2$ single-layer and bulk modes at $\Gamma$, together with their character, 
displacement direction, involved atoms, and frequency.

Overall, the single-layer and bulk
phonon dispersions have a remarkable resemblance. In the bulk, all single-layer
modes are split into two branches but since the inter-layer interaction
is weak, the splitting is very low (similar to the situation in graphite
and graphene.\cite{Wirtz2004} The only notable exception from this is the
splitting of the acoustic modes around $\Gamma$. In the single-layer,
the resulting low frequency optical modes are absent.

In the single layer, the high frequency $\Gamma$ modes
$E_{2g}^1$ and $E_{1u}$ collapse into the mode $E'$.
(From Fig.~\ref{modes_MoS2} it is evident that 
with increasing inter-layer distance, the modes
$E_{2g}^1$ and $E_{1u}$ acquire the same frequency.) 
Interestingly, as measured in Ref.~\onlinecite{Lee2010} and indicated
in Table~\ref{symmetry}, the bulk $E_{2g}$ mode is lower in 
frequency than the single-layer $E'$ mode. This contradicts the expectation
that the additional inter-layer interaction should increase the frequency
but is in line with the anomalous sign of the Davydov splitting between
the bulk $E_{2g}^1$ and $E_{1u}$ modes. The origin of this will be discussed
in section \ref{secthick}.
The out-of-plane mode $A_{1g}$ follows the expected trend that the 
inter-layer interaction increases the frequency with respect to   
the single-layer mode $A_1$.

The densities of states (DOS) of single-layer and bulk are represented 
in the right panels of Fig.~\ref{modes_MoS2}. 
The differences between single-layer and bulk DOS are minimal, 
except a little shoulder 
around 60 cm$^{-1}$ in the bulk DOS due to the low frequency optical modes. 
In both cases the highest
peaks are located close to the Raman active modes $E_{2g}^1$ and $A_{1g}$.

\subsection{WS$_2$}

\begin{figure}
\begin{center}
\includegraphics[width=7.6 cm]{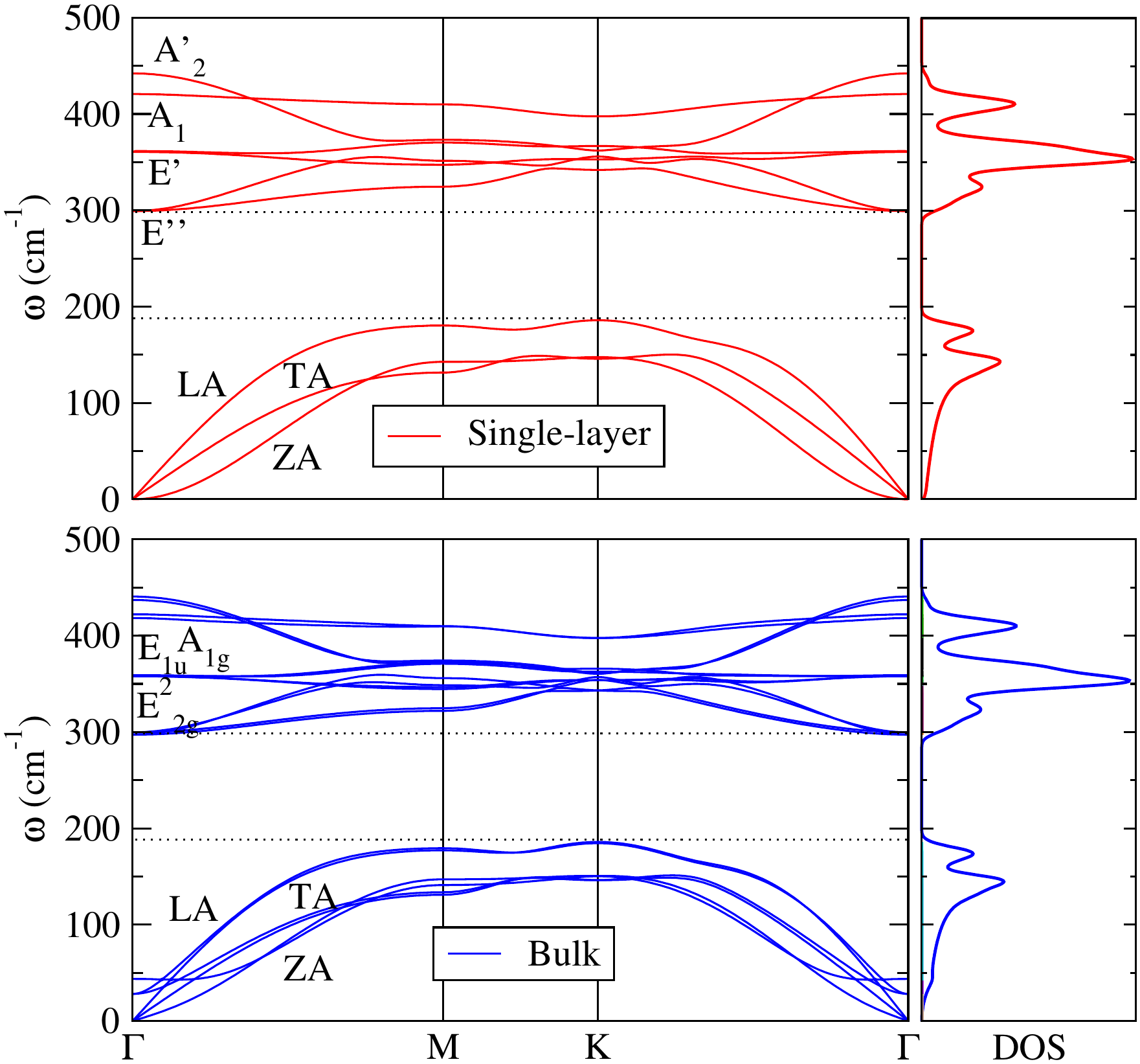}
\caption{Phonon dispersion curves and density of states of
1-layer and bulk WS$_2$.}
\label{ws2_phon}
\end{center}
\end{figure}

Figure~\ref{ws2_phon} shows the phonon dispersions of
single-layer and bulk WS$_2$, together with the density of states (DOS). 
The general features are identical to those of the dispersions of
MoS$_2$ (Fig.~\ref{mos2_phon}). The differences between single-layer and
bulk dispersions are similarly weak as in the case of MoS$_2$.
Thus, also the bulk DOS resembles very much that of the single-layer 
(except for the small shoulder $\sim 50$ cm$^{-1}$ due to the inter-layer
optical modes).

\begin{figure}
\begin{center}
\includegraphics[width=7.6 cm]{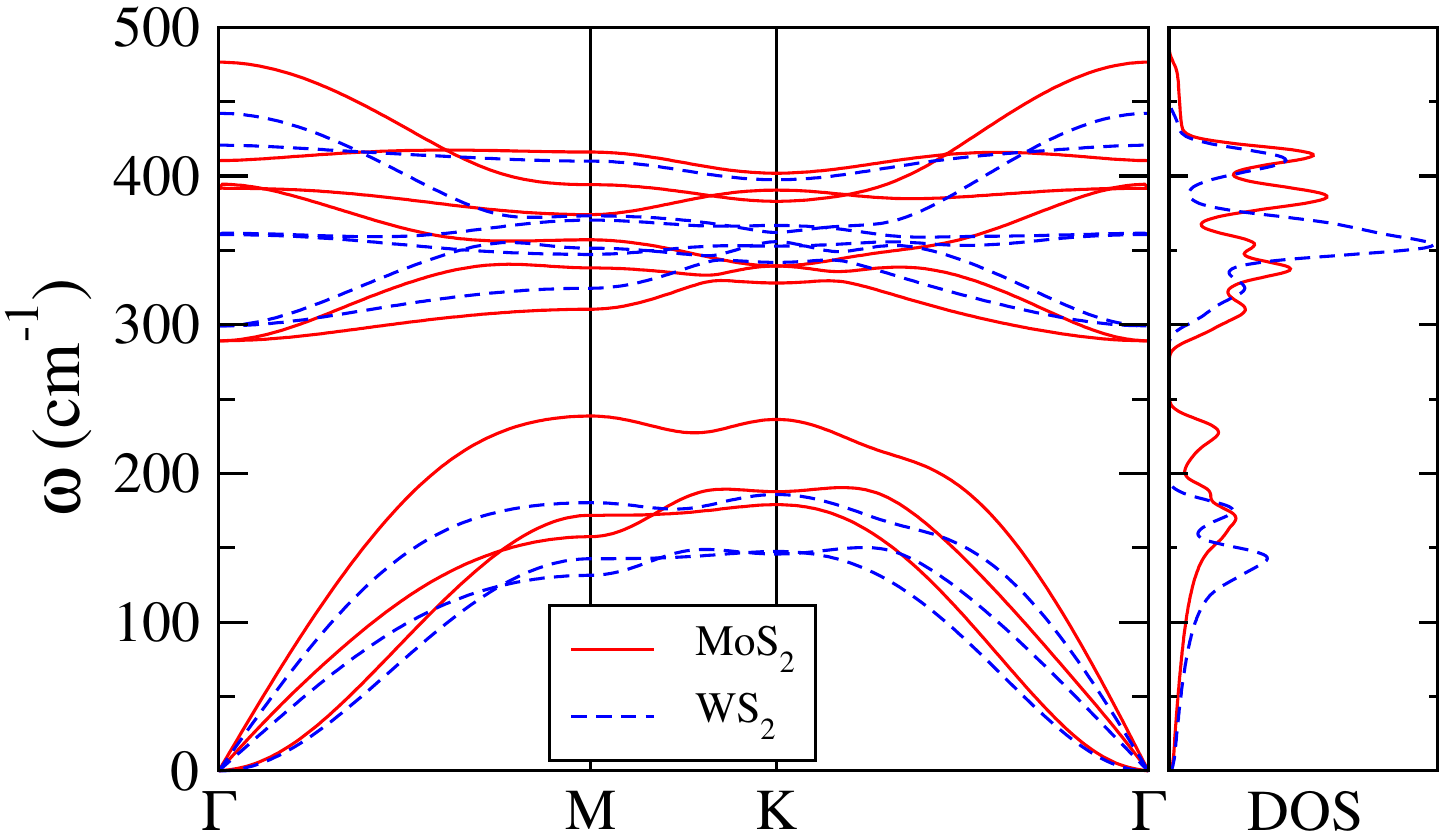}
\caption{Phonon dispersion curves for single-layer 
of MoS$_2$ (solid lines) and WS$_2$ (dashed lines). The density
of states is depicted in the right panel.}
\label{mos2_ws2}
\end{center}
\end{figure}

For a better comparison of MoS$_2$ and WS$_2$ single-layer
phonon frequencies, we have depicted them together in
Fig.~\ref{mos2_ws2}. In general, the WS$_2$ phonon bands are 
shifted down to lower frequencies with respect to the 
MoS$_2$ frequencies. The cause of this trend is the
larger mass of the tungsten atoms, and therefore their lower
vibration frequency (see Eq.~\ref{secular}). 
The only notable exceptions from this general downshift are
the branches associated to the mode $E''$, around 300 cm$^{-1}$
and to the $A_1$ mode around 410 cm$^{-1}$. In these modes, only
the sulfur atoms are vibrating (see Table~\ref{symmetry})
and thus their frequency is not affected by the mass of the 
metal atom (W or Mo), just by the strength of the covalent bond.

The larger mass of W leads to a larger frequency gap between low and high 
frequency modes (110 cm$^{-1}$) since the highest acoustic branch is 
pushed down. Furthermore, the difference between the modes $A_1$ and 
$E'$, is now of 60 cm$^{-1}$, three times larger than in the case
of MoS$_2$. 

The density of states of the WS$_2$ single-layer
is also appreciably different from that of MoS$_2$. 
While at low frequencies the DOS has
two well differenced peaks, as in Fig.~\ref{mos2_phon}, for higher energies 
one peak stands out from the others, at frequency of $\sim 350$ cm$^{-1}$, 
and associated mainly to the $\Gamma$-point mode $E'$. 

\section{Evolution of $A_{1g}$ and $E_{2g}^1$ phonon modes 
with the number of layers}
\label{secthick}

The understanding of the frequency trends of the 
$A_{1g}$ and $E_{2g}$ modes with varying layer thickness  
requires a more refined analysis. With the aim
of explaining the Raman scattering experiments of Ref.~\onlinecite{Lee2010}
we have calculated the phonon frequencies at the $\Gamma$ point
for single, double- and triple-layers and we discuss the evolution of
the atomic force constants from single-layer to bulk in detail.

\begin{figure}
\begin{center}
\includegraphics[width=7.6 cm]{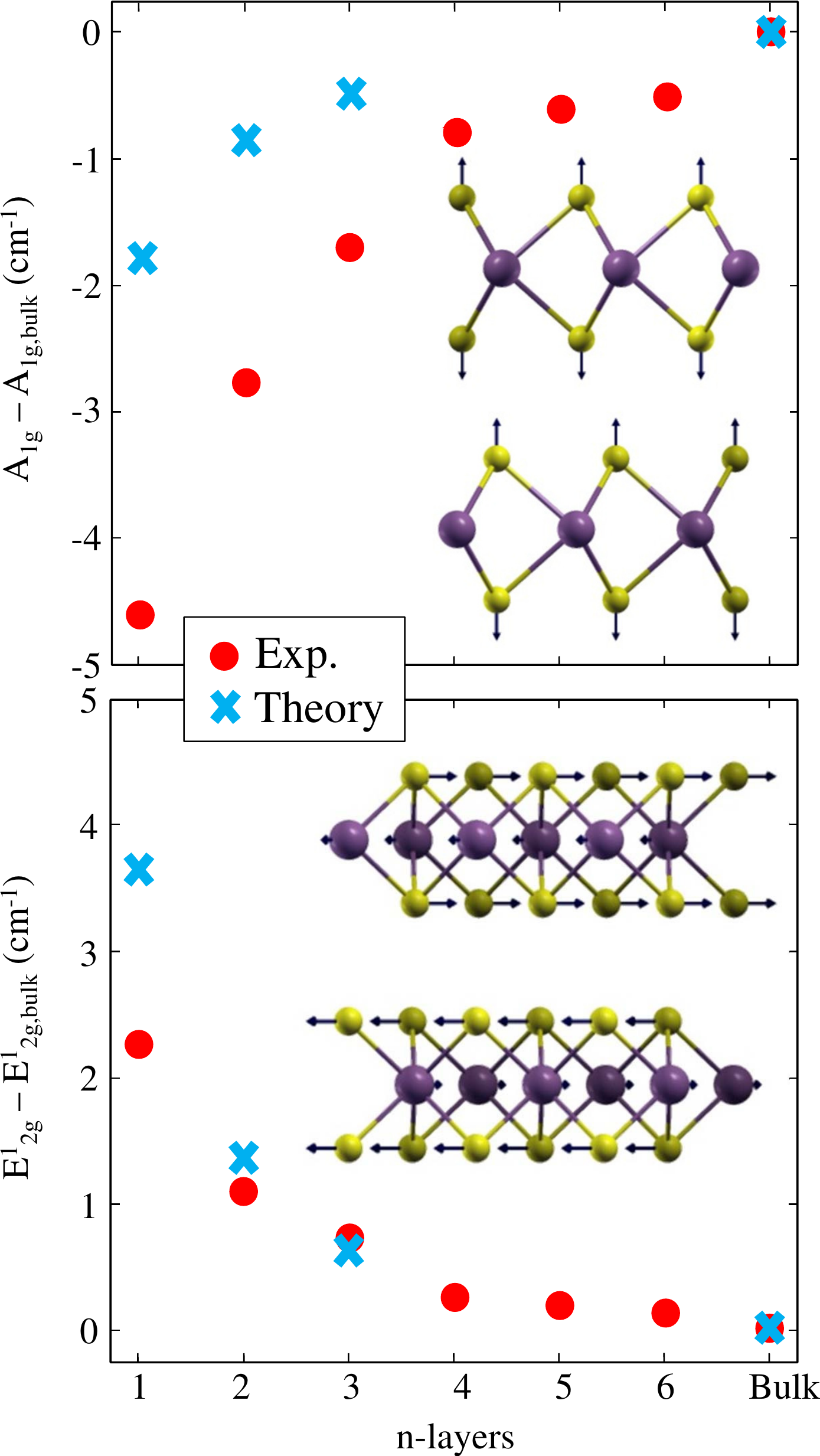}
\caption{Phonon frequencies of $A_{1g}$ and $E_{2g}^1$ modes 
as a function of
MoS$_2$ layer thickness, as obtained 
in this work (squares) and experimental data from
Ref.~\onlinecite{Lee2010} (circles). We plot the frequency differences
with respect to the corresponding bulk modes.
The insets represent the phonon modes at $\Gamma$ point.}
\label{phonon_layer}
\end{center}
\end{figure}

Figure~\ref{phonon_layer} shows the
phonon frequency of $A_{1g}$ and $E^1_{2g}$ modes 
as a function of the number of
layers. Since LDA tends to overestimate the phonon frequencies,
it is reasonable to represent the difference between the $n$-layers
frequency and the bulk frequency instead of comparing absolute theoretical
and experimental values. In such a representation we observe
that our calculation properly reproduce the up-shift of the $A_{1g}$ mode
and the down-shift of the $E^1_{2g}$ mode with increasing layer number.
The quantitative differences between theory and experiment are mainly due 
to the limited precision of the description of the inter-layer interaction 
by the LDA.

After we have shown that DFPT-LDA reproduces
the experimental trends we would like to give an explanation
of this behavior, taking advantage of the detailed knowledge of the
atomic force constants, available in \textit{ab initio} methods. 
We interpret the changes in phonon frequency through an analysis of the 
real-space force constants (Eq.~\ref{force}), in particular of the
self-interaction term (Eq.~\ref{selfforce}).
Going from the single-layer to a multi-layer, this term changes in two ways:
i) the short range term increases due to the (weak, but non-negligible) interaction 
with atoms from neighboring layers; (ii) the long-range Coulomb interaction
changes, because the sum extends over effective charges from all layers
and the effective screening of the Coulomb potential increases.

We start our analysis with the out-of-plane mode $A_{1g}$. From 
Fig.~\ref{modes_MoS2}, it is intuitively clear
that the interaction between sulfur atoms of neighboring layers can influence
substantially the frequency. Going from the single-layer to the double-layer,
one ``adds'' an additional ``spring'' between the atoms S and S' on neighboring
layers which leads to an increase of the $A_{1g}$ mode frequency with increasing
number of layers. As only sulfur atoms are involved in this mode, we just need to examine
the sulfur self-interaction term ($C_{Sz,Sz}$) and the atomic force constant ($C_{Sz,S'z}$) 
between nearest neighbors, that belong 
to adjacent layers (atoms joined by the spring in Fig.~\ref{modes_MoS2}). We have 
summarized in Table~\ref{forces} the results. Note that the small variation of 
the term $C_{Sz,Sz}$ from single-layer to bulk prevents it from being the main 
cause of a frequency increasing of almost 5 cm$^{-1}$. Thus, the term $C_{Sz,S'z}$ 
has a negative value which implies a binding force (``spring'') between
atoms S and S' which leads to an increase of frequency.
Force constants related to farer neighbors are negligible in comparison with $C_{Sz,S'z}$. This 
demonstrates that the weak interlayer interaction is the 
main cause of the frequency increasing with the number of layers.

One might expect that the same argument holds for the $E_{2g}^1$ mode: the
additional ``spring'' between sulfur atoms from neighboring layers should 
increase the frequency with increasing number of layers as well.
However, theoretical and experimental results show the opposite behavior. 
The reason can be found in the self-interaction terms $C_{Mox,Mox}$ for
Mo and S in Table~\ref{force}: while the difference between single-layer
and bulk is negligible for the sulfur atoms, one observes a considerable
decrease for the molybdenum atoms. Interestingly, the short range contribution
to the self-interaction increases (as one would expect from the inter-layer
interaction). However, the long-range Coulomb part\cite{Gonze1997} decreases 
considerably.
In the appendix, we show that this decrease is related to the strong
increase of the dielectric tensor (both in-plane and out-of-plane) when
going from the single-layer (represented in our calculations by a periodic 
stacking of single-layers with large vacuum between) to the bulk (see
Table~\ref{forces}). 
We note that one might associate the change in frequency to the 
differences in lattice constant and interatomic distances 
in bulk and single-layer, respectively. However, the 
small differences shown in Section II are not enough to account
for the observed magnitude of the $E_{2g}^1$ frequency change.\cite{contract}
  
Therefore, the decrease of the $E_{2g}^1$ phonon frequency is associated
to a stronger dielectric screening of the long range Coulomb interaction 
in few-layer and bulk 
MoS$_2$. The effect is particularly pronounced for the molybdenum atoms, 
as Table~\ref{forces} shows. Our
analysis also explains why previous empirical models 
have not been able to explain the experimental
observations, due to the difficulty to include this subtle change in 
the parameters.

\begin{table*}
\begin{tabular}{lc|cc|ccc|cc}
\hline
\hline
                           &Atom & $C_{Sz,Sz}$  &$C_{Sz,S'z}$ & $C_{Mox,Mox}$ &  Long-range  &  Short-range &  $\epsilon_{xx} = \epsilon_{yy}$  &  $\epsilon_{zz}$       \\
\hline
 \multirow{2}{*}{Bulk}     &Mo   & -         & -        &0.27660  & 0.00308       & 0.27352                   & \multirow{2}{*}{15.40}          & \multirow{2}{*}{7.43}\\
                           &S    & 0.15837   & -0.00099 &0.12310  & 0.00071       & 0.12239                   &                                 &                      \\
\hline
 \multirow{2}{*}{1-layer}  &Mo   & -         & -        &0.28275  & 0.01058       & 0.27217                   & \multirow{2}{*}{7.36}           &\multirow{2}{*}{1.63}\\
                           &S    & 0.15860   & -        &0.12320  & 0.00265       & 0.12055                   &                                 &\\
\hline\hline  
\end{tabular}
\caption{Self-interaction term $C_{S\alpha,S\alpha}$ and $C_{Mo\alpha,Mo\alpha}$ of the Mo and S atoms, and atomic force constants
between S atoms of adjacent layers, $C_{Sz,S'z}$,
as defined in Section~\ref{theory} (atomic units). The dielectric tensor $\bm{\epsilon}$ of both 
systems is also given (non-diagonal elements are zero).}
\label{forces}
\end{table*}

\section{Conclusions}

In conclusion, we have studied the phonon dispersions of MoS$_2$ and WS$_2$ single 
layers and bulk using density functional perturbation theory in the local
density approximation. We obtain good agreement with neutron diffraction
data as well as Raman and infrared absorption spectroscopies.
We have explored how the Raman active modes $A_{1g}$ and  
$E_{2g}^1$ change their frequencies when the number of layers varies, and 
confirm the recently reported down-shift of the $E_{2g}^1$ mode with
increasing number of layers. This unexpected behavior can be explained
by an increase of the dielectric screening which reduces the long range
Coulomb interaction between the effective charges and thus reduces the
overall restoring force on the atoms. We expect that this explanation also
holds for other polar layered materials where an anomalous Davydov splitting 
has been observed (such as GaSe\cite{Wieting1972} and GaS\cite{Kuroda1979}) 
or where the Raman peak shifts down with increasing number of layers
such as it has been recently observed for hexagonal BN.\cite{Gorbachev2011}

\begin{acknowledgements}
Funding was provided by the French National Research
Agency (ANR) through project bl-inter09\_482166.
Calculations were done at the IDRIS
supercomputing center, Orsay (Proj. No. 091827).
\end{acknowledgements}

\appendix
\section{}

The examination of the long-range atomic force constant formula can help 
to establish 
quantitatively its change with the variation of the dielectric tensor in the single-layer
and bulk MoS$_2$. However, the anisotropy of the 
crystalline structure of MoS$_2$ impedes a direct
relation between the long range atomic force constants and the dielectric tensor. The
long range contribution, $C_{I\alpha,J\beta}^{lr}$, can be written in terms of
the dielectric tensor $\bm{\epsilon}$, 
its inverse, $\bm{\epsilon}^{-1}$, the 
Born effective charges $Z^*_{I,\alpha\alpha'}$, and the 
interatomic distance $\bm{d}\equiv\bm{d}_{IJ}$, as defined in Ref.~\onlinecite{Gonze1997}:

\begin{align*}
C_{I\alpha,J\beta}^{lr} &= \sum_{\alpha',\beta'}Z^*_{I,\alpha,\alpha'}Z^*_{J,\beta,\beta'}
\left(\frac{(\epsilon^{-1})_{\alpha'\beta'}}{D^3} - 3\frac{\Delta_{\alpha'}\Delta{\beta'}}{D^5} \right) \\
&\times(det\epsilon)^{-1/2},
\end{align*}
where $\Delta_{\alpha}=\sum_{\beta}(\bm{\epsilon}^{-1})_{\alpha\beta}d_{\beta}$ is the 
conjugate of the vector $\bm{d}$, and the norm of the latter in this 
metrics is $D=\sqrt{\bm{\Delta}\cdot\bm{d}}$. This expression simplifies enormously by assuming 
diagonal the dielectric tensor, and 
$Z^*_{I,\alpha\alpha'}\equiv Z^*_{I,\alpha\alpha'}\delta_{\alpha\alpha'}$. After some algebra 
one obtains:

\begin{equation}
C_{I\alpha,J\beta}^{lr}=\frac{Z^*_{I,\alpha\alpha}Z^*_{J,\beta\beta}}{\sqrt{\epsilon_{xx}\epsilon_{yy}\epsilon_{zz}}}
\left( \frac{\epsilon^{-1}_{\alpha\beta}\delta_{\alpha\beta}}{D^3} - 3  
\frac{\epsilon^{-1}_{\alpha\alpha}\epsilon^{-1}_{\beta\beta}d_{\alpha}d_{\beta}}{D^5} \right).
\label{diel2}
\end{equation}

We can examine with an example how the long-range atomic force constants of single-layer and bulk MoS$_2$ are related with 
the dielectric tensors. Thus, we can evaluate the term $C_{I\alpha,J\beta}^{lr}$ for neighbor Mo atoms that belong to the same layer, 
with interatomic distance $\bm{d}=(d,0,0)$, and assuming the same distance
for single-layer and bulk. Thus, we obtain a simplified expression of Eq.~\ref{diel2}:

\begin{equation}
C_{Mo,x,Mo,x}^{lr}=-2\frac{(Z^*_{Mo,xx})^2}{\sqrt{\epsilon_{xx}\epsilon_{zz}}d^3}.
\end{equation}

The Born effective charges $Z^*_{Mo,xx}$ are almost equal for both systems, and using the dielectric tensors given in Table~\ref{forces}, 
we obtain:

\begin{equation}
\frac{C_{Mo,x,Mo,x}^{lr}(1l)}{C_{Mo,x,Mo,x}^{lr}(bulk)}=\sqrt{\frac{\epsilon_{xx,bulk}\epsilon_{zz,bulk}}{\epsilon_{xx,1l}\epsilon_{zz,1l}}}=3.09.
\label{diel3}
\end{equation}

From the {\it ab-initio} calculation of the atomic constants we obtain $C_{Mo,x,Mo,x}^{lr}(1l)/C_{Mo,x,Mo,x}^{lr}(bulk)=3.19$, which is in 
agreement with the value obtained in Eq.~\ref{diel3}. This demonstrates that
the difference in the long-range part of the force constants for single-layer and bulk originates from the different dielectric screening.

\bibliographystyle{/home/alejandro/Documentos/literature/tesis_prb}

\begin{thebibliography}{10}

\bibitem{Novoselov2005a}
K.~S. Novoselov, D.~Jiang, F.~Schedin, T.~J. Booth, V.~V. Khotkevich, S.~V.
  Morozov, and A.~K. Geim.
\newblock Proc. Natl. Acad. Sci. U.S.A. \textbf{102}, 10451 (2005).

\bibitem{Novoselov2004}
K.~S. Novoselov, A.~K. Geim, S.~V. Morozov, D. Jiang, Y. Zhang, S.~V. Dubonos,
I.~V. Grigorieva, A.~A. Firsov.
\newblock Science \textbf{306}, 666 (2004).

\bibitem{Wallace1947}
P.~R. Wallace.
\newblock Phys. Rev. \textbf{71}, 622 (1947).

\bibitem{Geim2007}
A.~K. Geim and K.~S. Novoselov.
\newblock Nat Mater \textbf{6}, 183 (2007).

\bibitem{son2006}
Y.-W. Son, M.L. Cohen, and S.G. Louie,
\newblock Phys. Rev. Lett. \textbf{97}, 216803 (2006).

\bibitem{giovannetti2007}
G. Giovannetti, P.A. Khomyakov, G. Brocks, P.J. Kelly, and J. van den Brink,
\newblock Phys. Rev. B \textbf{76}, 073103 (2007)

\bibitem{Zhang2009a}
Y.~Zhang, T.-T. Tang, C.~Girit, Z.~Hao, M.~C. Martin, A.~Zettl, M.~F. Crommie,
  Y.~R. Shen, and F.~Wang.
\newblock Nature \textbf{459}, 820 (2009).

\bibitem{Han2007} M.~Y.~Han, B.~Ozyilmaz, Y.~Zhang and P.~Kim, Phys. Rev. Lett.\textbf{98}, 206805 (2007).

\bibitem{Splendiani2010}
A.~Splendiani, L.~Sun, Y.~Zhang, T.~Li, J.~Kim, C.-Y. Chim, G.~Galli, and
  F.~Wang.
\newblock Nano Lett. \textbf{10}, 1271 (2010).

\bibitem{Mak2010}
K.~F. Mak, C.~Lee, J.~Hone, J.~Shan, and T.~F. Heinz.
\newblock Phys. Rev. Lett. \textbf{105}, 136805 (2010).

\bibitem{RadisavljevicB.2011}
B.~Radisavljevic, A.~Radenovic, J.~Brivio, V.~Giacometti, and A.~Kis.
\newblock Nat Nano \textbf{6}, 147 (2011).

\bibitem{Coleman2011}
J.~N. Coleman, M.~Lotya, A.~ONeill, S.~D. Bergin, P.~J. King, U.~Khan,
  K.~Young, A.~Gaucher, S.~De, R.~J. Smith, I.~V. Shvets, S.~K. Arora,
  G.~Stanton, H.-Y. Kim, K.~Lee, G.~T. Kim, G.~S. Duesberg, T.~Hallam, J.~J.
  Boland, J.~J. Wang, J.~F. Donegan, J.~C. Grunlan, G.~Moriarty, A.~Shmeliov,
  R.~J. Nicholls, J.~M. Perkins, E.~M. Grieveson, K.~Theuwissen, D.~W. McComb,
  P.~D. Nellist, and V.~Nicolosi.
\newblock Science \textbf{331}, 568 (2011).

\bibitem{Krivanek2010}
O.~L. Krivanek, M.~F. Chisholm, V.~Nicolosi, T.~J. Pennycook, G.~J. Corbin,
  N.~Dellby, M.~F. Murfitt, C.~S. Own, Z.~S. Szilagyi, M.~P. Oxley, S.~T.
  Pantelides, and S.~J. Pennycook.
\newblock Nature \textbf{464}, 571 (2010).

\bibitem{Lee2010}
C.~Lee, H.~Yan, L.~E. Brus, T.~F. Heinz, J.~Hone, and S.~Ryu.
\newblock ACS Nano \textbf{4}, 2695 (2010).

\bibitem{Wieting1971}
T.~J. Wieting and J.~L. Verble.
\newblock Phys. Rev. B \textbf{3}, 4286 (1971).

\bibitem{Ghosh1983}
P.~N. Ghosh and C.~R. Maiti.
\newblock Phys. Rev. B \textbf{28}, 2237 (1983).

\bibitem{Ataca2011}
C. Ataca, H. Sahin, E. Akt\"urk, and S. Ciraci,
\newblock J. Phys. Chem. C \textbf{115}, 3934 (2011).

\bibitem{Wakabayashi1975}
N.~Wakabayashi, H.~G. Smith, and R.~M. Nicklow.
\newblock Phys. Rev. B \textbf{12}, 659 (1975).

\bibitem{Livneh2010}
T.~Livneh and E.~Sterer.
\newblock Phys. Rev. B \textbf{81}, 195209 (2010).

\bibitem{abinit}
X. Gonze, J.-M. Beuken, R. Caracas, F. Detraux, M. Fuchs, G.-M. Rignanese,
L. Sindic, M. Verstraete, G. Z\'erah, F. Jollet, M. Torrent, A. Roy,
M. Mikami, Ph. Ghosez, J.-Y. Raty, D.C. Allan,
Comp. Mat. Sci. {\bf 25}, 478 (2002). The {\tt ABINIT} code results
from a common project of the Universit\'e Catholique de Louvain,
Corning Incorporated, and other collaborators (http://www.abinit.org).

\bibitem{Kohn1965}
W.~Kohn and L.~J. Sham.
\newblock Phys. {R}ev. \textbf{140}, A1133 (1965).

\bibitem{hgh}
C. Hartwigsen, S. Goedecker, J. Hutter,
\newblock Phys. Rev. B \textbf{58}, 3641 (1998).

\bibitem{Schutte1987} W.~J.~Schutte, J.~L.~de Boer, and F.~Jellinek, 
\newblock J. Solid State Chem.\textbf{70}, 207 ͑(1987͒).

\bibitem{burke} J. P. Perdew, K. Burke, and M. Ernzerhof, Phys. Rev. Lett.
\textbf{77}, 3865 (1996).

\bibitem{kresse}
G. Kresse, J. Furthm\"uller, and J. Hafner, 
\newblock Europhys. Lett. \textbf{32}, 729 (1995).

\bibitem{kern}
G. Kern, G. Kresse, and J. Hafner, 
\newblock Phys. Rev. B \textbf{59}, 8551 (1999).

\bibitem{serrano}
J. Serrano, A. Bosak, R. Arenal, M. Krisch, K. Watanabe, T. Taniguchi,
H. Kanda, A. Rubio, and L. Wirtz,
\newblock Phys. Rev. Lett. \textbf{98}, 095503 (2007).

\bibitem{allard}
A. Allard and L. Wirtz,
\newblock Nano Lett. \textbf{10}, 4335 (2010).

\bibitem{Marini2006}
A. Marini, P. Garcia-Gonzalez, A. Rubio, 
\newblock Phys. Rev. Lett. \textbf{96} 136404 (2006).

\bibitem{lebegue}
S. Leb\`egue, J. Harl, Tim Gould, J. G. Angy\'an, G. Kresse, and J. F. Dobson,
\newblock Phys. Rev. Lett. \textbf{105}, 196401 (2010).

\bibitem{Rydberg2003}
H.~Rydberg, M.~Dion, N.~Jacobson, E.~Schr\"{o}der, P.~Hyldgaard, S.~I. Simak,
  D.~C. Langreth, and B.~I. Lundqvist.
\newblock Phys. Rev. Lett. \textbf{91}, 126402 (2003).

\bibitem{Bruesch1982}
P.~Br\"{u}esch.
\newblock \emph{Phonons: Theory and Experiments I} (Springer-Verlag, 1982).

\bibitem{Gonze1997}
X.~Gonze and C.~Lee.
\newblock Phys. Rev. B \textbf{55}, 10355 (1997).

\bibitem{Baroni2001}
S.~Baroni, S.~de~Gironcoli, A.~Dal~Corso, and P.~Giannozzi.
\newblock Rev. Mod. Phys. \textbf{73}, 515 (2001).

\bibitem{CardonaBook}
M.~Cardona and P.~Y. Yu.
\newblock \emph{Fundamentals of {S}emiconductors} (Springer-Verlag, 1996).

\bibitem{Giannozzi1991}
P.~Giannozzi, S.~de~Gironcoli, P.~Pavone, and S.~Baroni.
\newblock Phys. Rev. B \textbf{43}, 7231 (1991).

\bibitem{Saito1998}
R.~Saito, G.~Dresselhaus, and M.~S. Dresselhaus.
\newblock \emph{Physical Properties of Carbon Nanotubes} (Imperial College
  Press, London, 1998).

\bibitem{Wirtz2004}
L.~Wirtz and A.~Rubio.
\newblock Solid State Commun. \textbf{131}, 141 (2004).

\bibitem{contract} 
In order to proof this hypothesis, we have also examined
the frequencies of a single-layer setting bulk interatomic distances, in order
to mimic as much as possible the bulk environment onto the single layer. The 
$E_{2g}^1$ is now 390.3 cm$^{-1}$, which is still larger than the bulk, with a 
value of 387.8 cm$^{-1}$. Therefore, we can neglect the distinct interatomic distances
as the main cause of the $E_{2g}^1$ frequency trend.

\bibitem{Wieting1972}
T.J. Wieting and J.L. Verble,
\newblock Phys. Rev. B \textbf{5}, 1473 (1972).

\bibitem{Kuroda1979}
N. Kuroda and Y. Nishina,
\newblock Phys. Rev. B \textbf{19}, 1312 (1979).

\bibitem{Gorbachev2011}
R.V. Gorbachev, I. Riaz, R.R. Nair, R. Jalil, L. Britnell,
B.D. Belle, E.W. Hill, K.S. Novoselov, K. Watanabe, T. Taniguchi, 
A.K. Geim, and P. Blake,
\newblock Small \textbf{7}, 465 (2011).
\end{thebibliography}

\end{document}